\documentclass[a4paper]{article}
\usepackage{amsmath,amssymb,fullpage}
\usepackage{color,ulem,graphicx}

\begin{document}

%
%

\title{Pre-seismic ionospheric anomalies detected before the 2016 Taiwan
  earthquake}
%
%

\author{   Shin-itiro Goto$^{1}$, 
  Ryoma Uchida$^{1}$, 
   Kiyoshi Igarashi$^{1}$, 
   Chia-Hung Chen$^{2}$,\\
   Minghui Kao$^{1}$, 
 and Ken Umeno$^{1}$\\ 
1 Department of Applied Mathematics and Physics,\\
Graduate School of Informatics, 
Kyoto  University, Kyoto, Japan.\\
2 Department of Earth Sciences, National Cheng Kung University,
Tainan, Taiwan.
}

\maketitle




%
%

\begin{abstract}

On Feb. 5 2016 (UTC), an earthquake with moment magnitude 6.4 occurred
in southern Taiwan, known as the 2016 (Southern) Taiwan earthquake.
In this study,    
evidences of seismic earthquake precursors for this 
earthquake event are investigated.
Results show that ionospheric anomalies in Total Electric
Content (TEC) can be observed before the earthquake. 
These anomalies were obtained by processing TEC data,
where such TEC data 
are calculated from phase delays of signals observed at 
densely arranged ground-based stations in Taiwan 
for Global Navigation Satellite Systems. 
This shows that 
such anomalies were detected within 1 hour
before the event.

\end{abstract}

%
%

%



%
%
%

\section{Introduction}
\label{sec:intro}

The ionosphere is an ionized medium which  
can affect the radio communications. 
The electron density in the ionosphere 
is disturbed by various phenomena such as solar
flares \cite{Donnelly1976}, 
volcanic eruptions \cite{Igarashi1994},
flying objects \cite{Mendillo1975},
earthquakes \cite{Ogawa2012}, and so on.  
These electron density 
disturbances are observed with
Total Electron Contents (TECs) at 
ground-based Global Navigation Satellite Systems (GNSS) receivers.
With GNSS that can monitor variations of TEC, it has been reported 
\cite{Heki2011,H-Enomoto2015}
that pre-seismic ionospheric electron density anomalies appeared frequently
before large earthquakes, which could be
caused by the earthquake-induced electromagnetic process before such 
earthquakes. 
Furthermore, such TEC anomalies were found in the 2016
Kuramoto earthquake (Mw7.3) \cite{IU2017}.

Taiwan is located in an active 
seismic area \cite{Liu2000GRL,Oyama2008} and has ground-based stations for 
GNSS with densely arranged receivers. Thus, Taiwan is a  suitable region
to explore relations 
between earthquakes 
and pre-seismic ionospheric TEC anomalies.
 Also, Taiwan has been focused from a viewpoint of
  electric currents in the earth crust and electromagnetic emissions
  preceding to earthquakes \cite{Freund2003}.
An  earthquake occurred around southern Taiwan in Feb. 2016 
(19:57 UT, 5 Feb. 2016, Mw6.4, depth $23.0$ km, $22.94$ N $120.6$ E  ).
As reported in U.S. Geological Survey \cite{USGSReport2016Taiwan},
this event was the result of oblique thrust faulting at shallow-mid
crustal depths.
Rupture occurred on a fault oriented either northwest-southwest,
and dipping shallowly to the northeast,
or on a north-south striking structure dipping steeply to the west.  
At the location of the earthquake, 2 tectonic plates
converge in a northwest-southeast direction at a velocity of
about $80$ mm a year.

\begin{figure}[ht!]
\includegraphics[height=4.3cm]{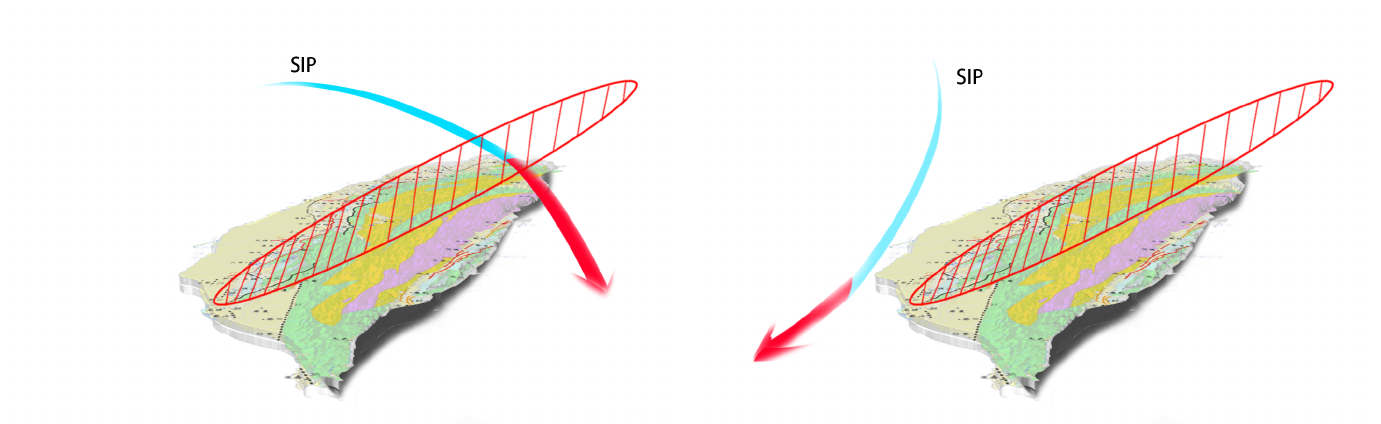}
  \caption{ (left) When a Sub-Ionospheric Point (SIP) track
  crosses the projection of an assumed ionospheric anomaly area (shaded in red).
  (right) When a SIP track does not cross the projection of the anomaly area.
}
\label{Figures-AB}
\end{figure}

In this paper pre-seismic ionospheric TEC anomalies
for the 2016 Taiwan earthquake (Mw6.4) are shown  
with the method described in \cite{IU2016}.
 This analysis is 
referred to as CoRelation Analysis (CRA) in this paper.  
With CRA, 
the TEC anomalies were detected for 
the Kuramoto earthquake \cite{IU2017}, where this earthquake 
is classified as an intraplate earthquake 
and its moment 
magnitude is less than Mw 8.0. 
It is then of interest to explore 
if TEC anomaly is observed with this method for 
intraplate earthquakes whose 
moment magnitudes are less than Mw 7.0.  
After TEC anomalies will be shown in this paper, a possible    
link connecting satellite tracks with an assumed ionospheric
anomaly area over Taiwan will be argued. This is to investigate why such
anomalies can be observed with some specific satellite tracks
( See Figure \ref{Figures-AB}).

\vspace*{1cm}

\section{Method}
\label{section-correlation-analysis-explanation}
The TEC along the line of sight, called slant TEC, can be
calculated with the phase delay of signals sent from GNSS satellites 
observed at ground-based GNSS receivers. In the following slant TEC 
is abbreviated as TEC, and TEC data are 
calculated from signals sent from
the Global Positioning System (GPS, American GNSS) satellites.  Also,  
the standard unit for TEC,
TECU that is $10^{16}$ el/$m^2$, is used. 
For simplifying various calculations,  
electrons in the ionosphere are approximated to be in a thin layer, and 
this layer for Taiwanese GNSS stations is set to $325$ km
above the ground.
At this altitude,
the electron density is the maximum in the ionosphere. 
The intersection of the line of sight from a GNSS 
receiver with this thin layer is called 
Ionospheric Piercing Point (IPP), and
its projection onto the surface of the earth is called
Sub-Ionospheric Point (SIP). SIP tracks
for Taiwanese GNSS stations  shown in this paper were calculated
with this altitude of the thin layer.

For large earthquakes the CoRrelation Analysis (CRA) was shown to  
 detect earthquake precursors \cite{IU2016}. 
  The analysis, CRA,  is based on calculating correlations among TEC values,
  where how to calculate these correlations is summarized as follows. 

\begin{enumerate}
\item
  Choose a central GNSS station, a satellite,
  and 3 parameters denoted by 
  $t_{\,\mathrm{sample}}$, $t_{\,\mathrm{test}}$, and $M$.
  The parameter $t_{\,\mathrm{sample}}$
  is the time-length of data used for obtaining a regression curve, 
  $t_{\,\mathrm{test}}$ the time-length for testing the difference between
  the regression curve
  and the obtained data, and $M$ the number of GNSS stations located 
  around the central station for the correlation analysis. 
  We fix $t_{\,\mathrm{sample}}$ and $t_{\,\mathrm{test}}$ to be $2$ and $0.25$
  h    throughout this paper.
\item
  Let SampleData be the TEC data for the time-duration from $t$
  to $t+t_{\,\mathrm{sample}}$ for each station labeled by $i$.
  Also, let TestData be the TEC data for the time-duration
  from $t+t_{\,\mathrm{sample}}$ to $t+t_{\,\mathrm{sample}}+t_{\,\mathrm{test}}$.
\item
  Fit a curve to SampleData by the least square method
  with a polynomial curve whose degrees is 7.  
  
\item
  Calculate a deviation of TestData from the regression curve
  for the time-duration $t_{\,\mathrm{test}}$, and such a deviation
  is denoted by $x_{i,t'}$ where $t'$ is such that
  $t_{\,\mathrm{sample}}\leq t'\leq t_{\,\mathrm{sample}}+t_{\,\mathrm{test}}$.
\item  
  Calculate the correlation defined as
  $$
  C(T)=\frac{1}{NM}\sum_{i=1}^{M}\sum_{j=0}^{N-1}
  x_{\,i,t+t_{\mathrm{sample}}+j\Delta t}\,x_{\,0,t+t_{\mathrm{sample}}+j\Delta t},
  $$
  where $T=t+t_{\,\mathrm{sample}}+t_{\,\mathrm{test}}$, 
  $N$ is the number of data points for TestData,
  $\Delta\, t=30$ sec
  the sampling interval for TestData  
  so that $\Delta\,t=t_{\,\mathrm{test}}/(N-1)$, and 
  $i=0$ the special label indicating the central GNSS station.
 The physical dimension of $C(T)$ is the square of TECU. 
\end{enumerate}
In what follows the word ``anomaly'' is used, when the correlation value
is high  enough. To quantify and define abnormality,
careful discussions are needed. 
Thus,  
we compare correlation values in view of days, SIP tracks, 
and distances among
GNSS stations. The details of these are discussed in
Section\,\ref{section-data-presentation-2016}.

\section{Data Presentation}
\label{section-data-presentation-2016}
In this section 
(i) raw TEC data and (ii) correlations are shown 
  with respect to GNSS ground-based stations, and with respect to
  GPS satellites 17 and 28.
A variety of choices for stations and satellites enables one to argue   
how TEC anomalies can be observed with respect to the SIP track.
For this reason GNSS stations were chosen from the south, middle,
and north of Taiwan.

TEC data
were obtained from the original GPS data 
with the method in \cite{Otsuka2002}, and the original 
GPS data were obtained from the   
Central Weather Bureau (CWB) in Taiwan.

\subsection{Results for the event day}
  In this subsection results for the event day are shown.  
The parameter $M$ was chosen to be $1$, which  
is equivalent to $2$ GNSS stations for CRA.

\subsubsection{Stations in the south of Taiwan}

\begin{figure}[ht!]
\includegraphics[height=14.0cm]{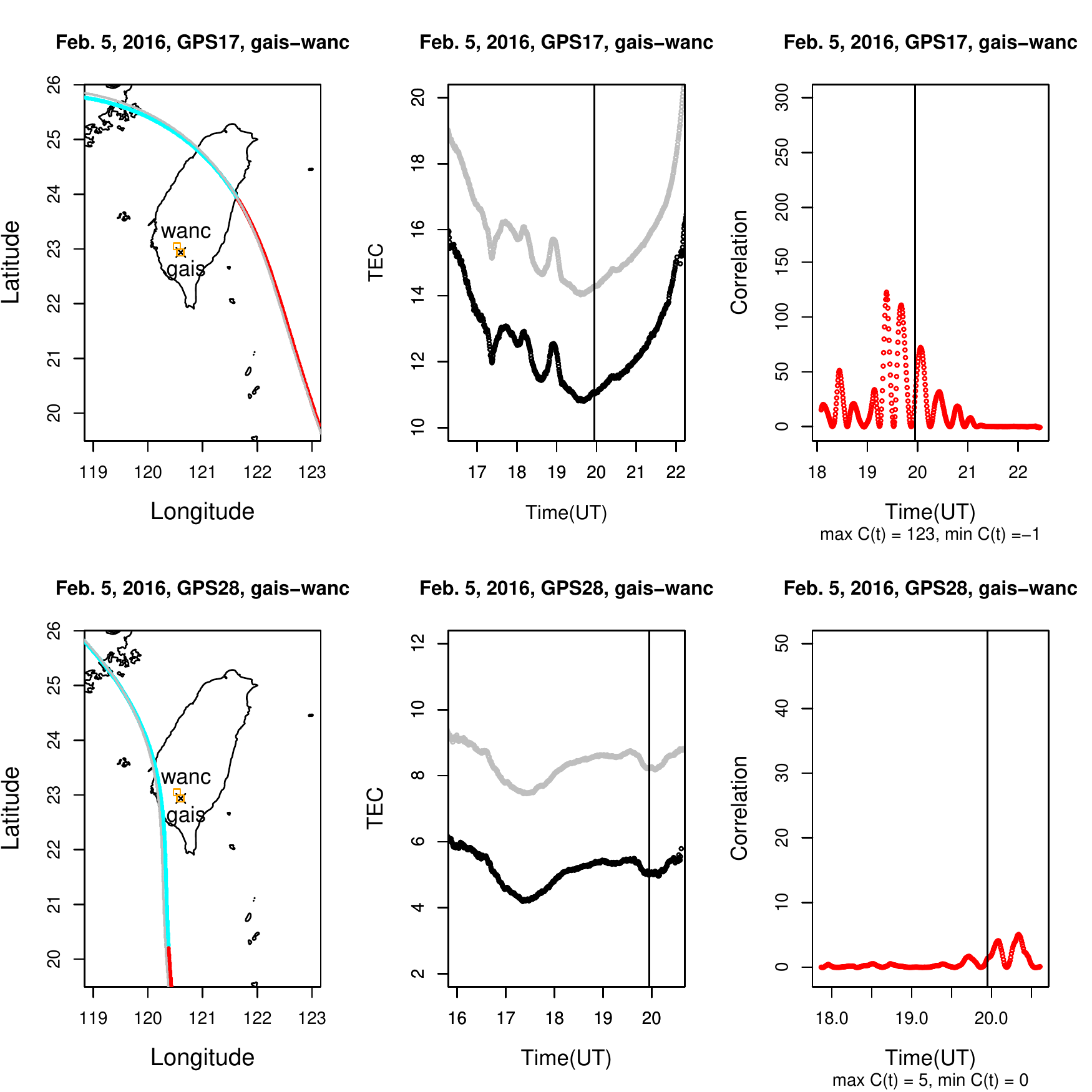}  
\caption{
  (left) SIP tracks for GPS satellites 17 and 28.
  Red: Track for ``gais'' after the event,
  Cyan: Track for ``gais'' before the event,
  Gray: Track for ``wanc'',
  $\times$: Epicenter. 
  (middle) Time series of TEC,
  the vertical lines indicate the time (UT) when the event occurred.
  Black: TEC obtained at ``gais'', 
  Gray: TEC obtained at ``wanc'',
  where the TEC values were shifted by hand for the guide of eyes. 
  (right) Time series of correlations obtained with CRA.
  }
\label{figure-south-event-day-M1}
\end{figure}

Figure \ref{figure-south-event-day-M1} shows results for stations located in
the south of Taiwan. 
 We chose ``gais'' as the central station, since it is 
the nearest station from the epicenter.
The distance is about $1.0$ km. 
The other station for CRA is ``wanc'', and it is the
next nearest station for which TEC data were available.  
The distance between ``wanc'' and the epicenter
is about $14.1$ km.
From the left panels, the SIP tracks for
  GPS satellite 17 crossed the projection of the
  assumed ionospheric anomaly area, and 
  most of the SIP tracks for GPS satellite 28 did not cross it.
  ( See figure \ref{Figures-AB} ).  
From the middle panels, 
it is verified that the obtained $2$ time-series 
of TEC data  were similar to each other. 
It is reasonable, 
since these 2 stations are geographically close.
Time-series of correlations  were obtained by processing 
this kind of TEC data. 
As it can be read off
from the right upper panel showing correlations obtained with CRA
  for GPS satellite 17,   
about 40 min before the event,   
a period of time was observed 
  where correlation values  
are high. 
Its maximum value was $123$,
which was observed at 19:23 UT. 
Besides, as can be seen from the right panels,
the correlation values for GPS satellite 28 were lower than those of 17. 
This comparison between the time-series of correlations   
indicates that how TEC anomalies can be observed is with respect to
the SIP track 
( See Section \ref{section-discussion} for discussions ). 

In the following, to explore a spatial dependency of TEC anomalies, we show   
TEC data and their correlations with CRA 
obtained at stations located other areas in Taiwan.

\subsubsection{Stations in the middle of Taiwan}
\begin{figure}[ht!]
\includegraphics[height=14.0cm]{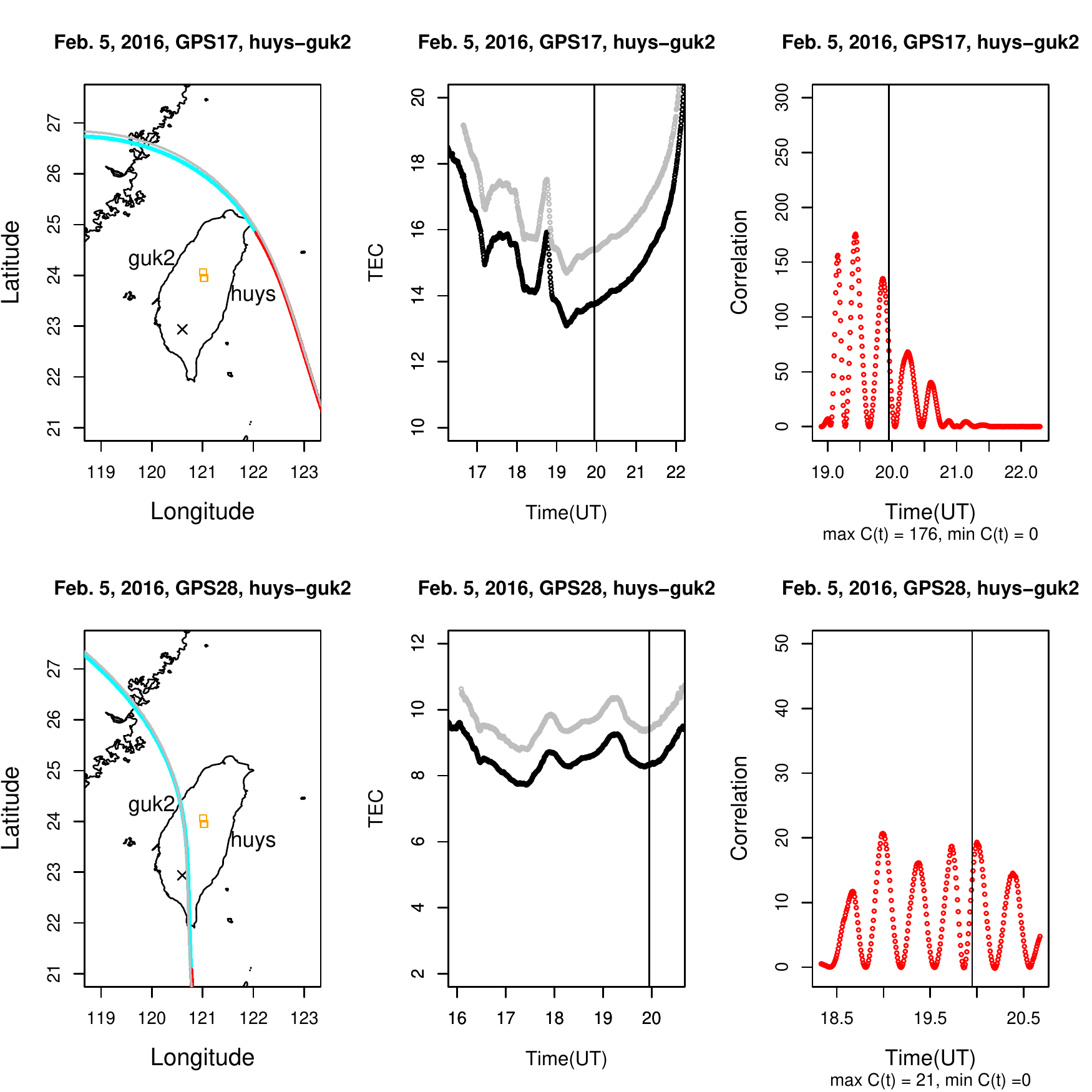}
\caption{
  (left) SIP tracks for GPS satellites 17 and 28.
  Red: Track for ``huys'' after the event,
  Cyan: Track for ``huys'' before the event,
  Gray: Track for ``guk2'',
  $\times$: Epicenter. 
  (middle) Time series of TEC,  
  Black: TEC obtained at ``huys'',  
  Gray: TEC obtained at ``guk2'', where the TEC values were shifted by hand for the guide of eyes. 
  (right) Time series of correlations obtained with CRA.
} 
\label{figure-middle-event-day-M1}
\end{figure}

Figure \ref{figure-middle-event-day-M1} shows results for stations located in
the middle of Taiwan.
We chose ``huys'' as the central station for CRA, 
     and the other station was ``guk2''.  
The distance between these stations is about 
$12.4$ km, and that between the epicenter and ``huys'' is about $120.2$ km. 
From the left panels, SIP tracks for GPS satellite 28
crossed the projection of the anomaly area,
which was contrary to the case of the south of Taiwan.
The SIP tracks for GPS satellite 28 was north-south, which was different
from the case for 17. 
It can be seen from the right panels, 
as well as the case of the south stations, 
correlation values associated with GPS satellite 17
were higher than those with 28.
The maximum correlation value associated with the satellite 17 was 176,
which was observed at 19:26 UT. 
Notice that the correlation values associated with GPS satellites 17 and 28 
became higher than those for the south stations. 

In the following, to see more details of this spatial dependency,
we focus on TEC data and their correlations obtained with CRA 
for stations located in the north of Taiwan. 

\subsubsection{ Stations in the north of Taiwan}
\begin{figure}[ht!]
\includegraphics[height=14.0cm]{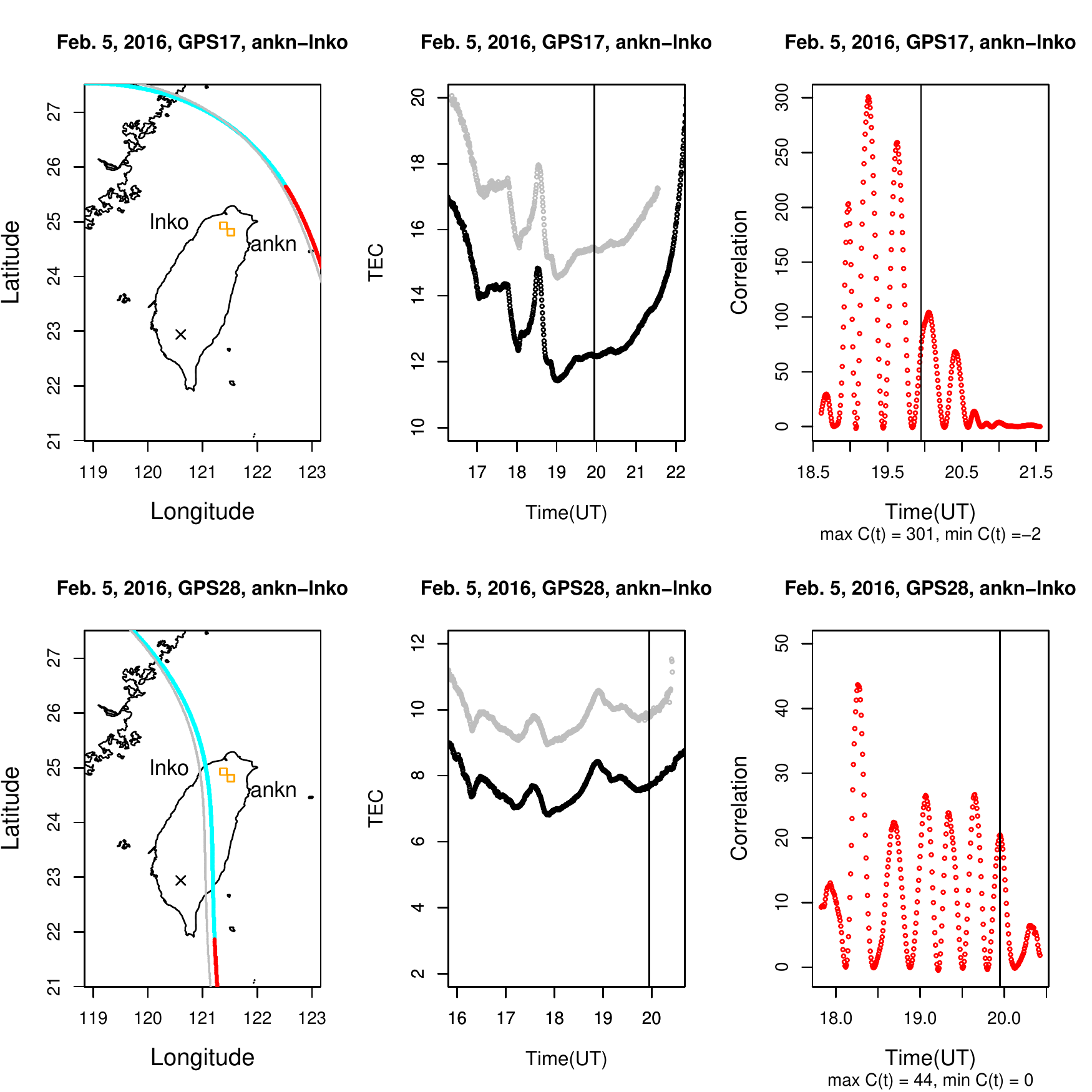}
\caption{
  (left) SIP tracks for GPS satellites 17 and 28.
  Red: Track for ``ankn'' after the event,
  Cyan: Track for ``ankn'' before the event,
  Gray: Track for ``lnko'',
  $\times$: Epicenter. 
  (middle) Time series of TEC,  
  Black: TEC obtained at ``ankn'',  
  Gray: TEC obtained at ``lnko'', where the
  TEC values were shifted by hand for the guide of eyes. 
  (right) Time series of correlations obtained with CRA.
} 
\label{figure-north-event-day-M1}
\end{figure}

Figure \ref{figure-north-event-day-M1} shows results for stations located in
the north of Taiwan. 
We chose  ``ankn'' as the central station for CRA,
and the other station was ``lnko''.  
The distance between these stations is about 
$19.4$ km, and that between the epicenter and ``ankn'' is about
$227.3$ km.
From the left panels, the SIP tracks for GPS satellites 17 and 28
crossed the projection of the assumed anomaly area. 
The maximum correlation value associated with the satellite 17 was 301,
which was observed at 19:14 UT.  
Notice from the right panels
that the correlation values associated with GPS satellites 17 and 28 
were higher than those for the stations located in the south and middle.  
 
In the next subsection the details of TEC anomalies are argued.
In particular, how correlations depend on $M$, whether or not
TEC anomalies were observed before the event day, and so on, are
focused. 

\subsection{Detailed results for the south stations}
In the previous subsection, TEC anomalies were found about 40 min before
the event on Feb. 5, where $M=1$ was used for CRA.  
To see if such anomalies occurred days before the event or not, we show
TEC data and their correlations with particular focus on the south stations. 
For the same purpose, $M$ dependency of correlations are shown as well. 
\subsubsection{TEC data for days before the event}

\begin{figure}[ht!]
\centerline{\includegraphics[height=15.0cm]{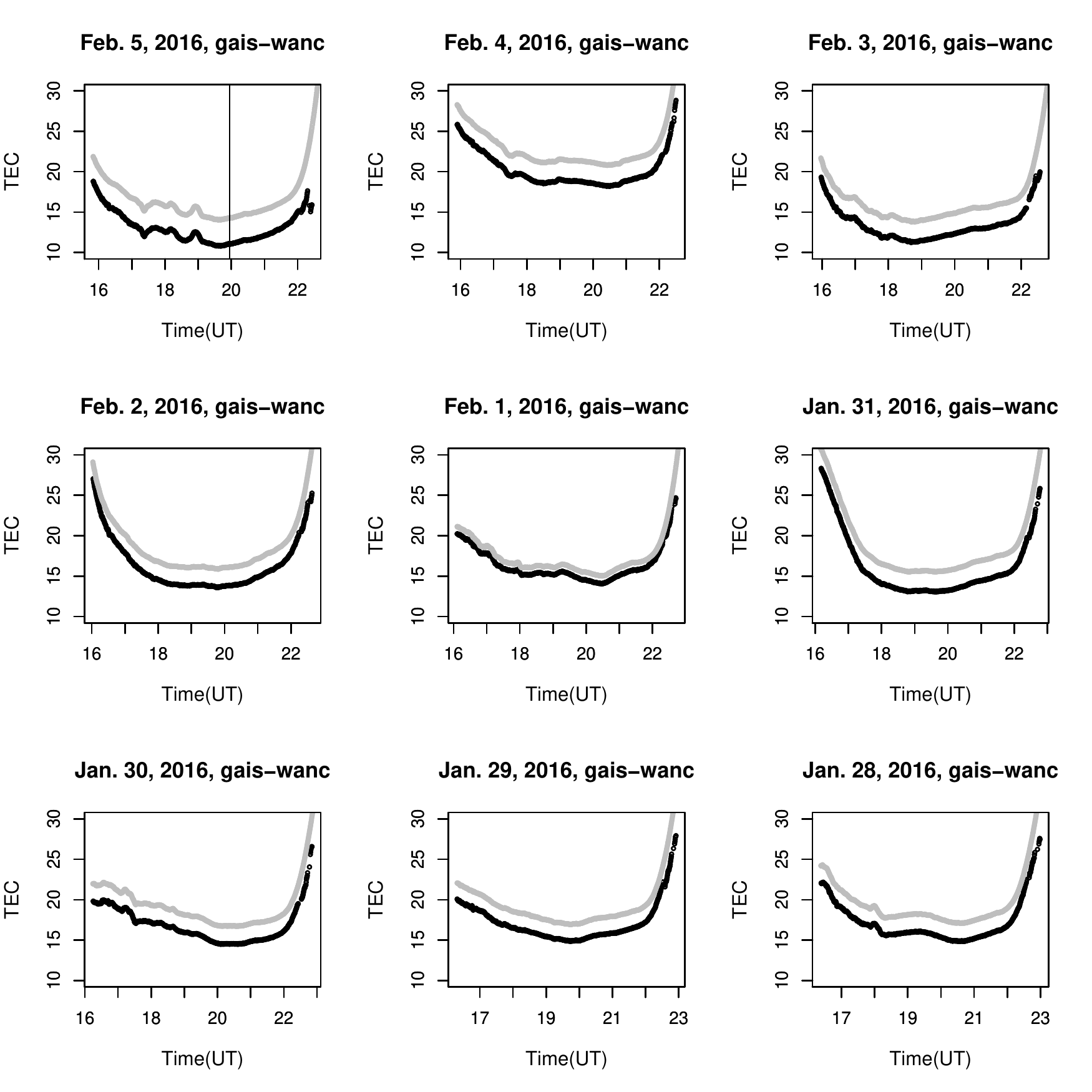}}
\caption{Time-series of TEC data observed
at  the stations ``gais'' (black) and ``wanc'' (gray) 
for the period from Jan. 28 to Feb. 5 in 2016, where the TEC values obtained
at ``wanc'' were shifted by hand for the guide of eyes.  
  GPS satellite 17 was used, and the vertical line on Feb. 5 indicates
  the time when the event occurred.
}
\label{figure-tainan-gais-wanc-raw-TEC-9}
\end{figure}

Figure \ref{figure-tainan-gais-wanc-raw-TEC-9} 
shows the time-series of TEC data observed for the consecutive 9 days   
including the event day.
These time-series were obtained
with GPS satellite 17 associated with the stations ``gais'' and ``wanc''.
From this figure, it is verified that the obtained $2$ time-series on each day
 were similar.

\begin{figure}[ht!]
\centerline{\includegraphics[height=15.0cm]{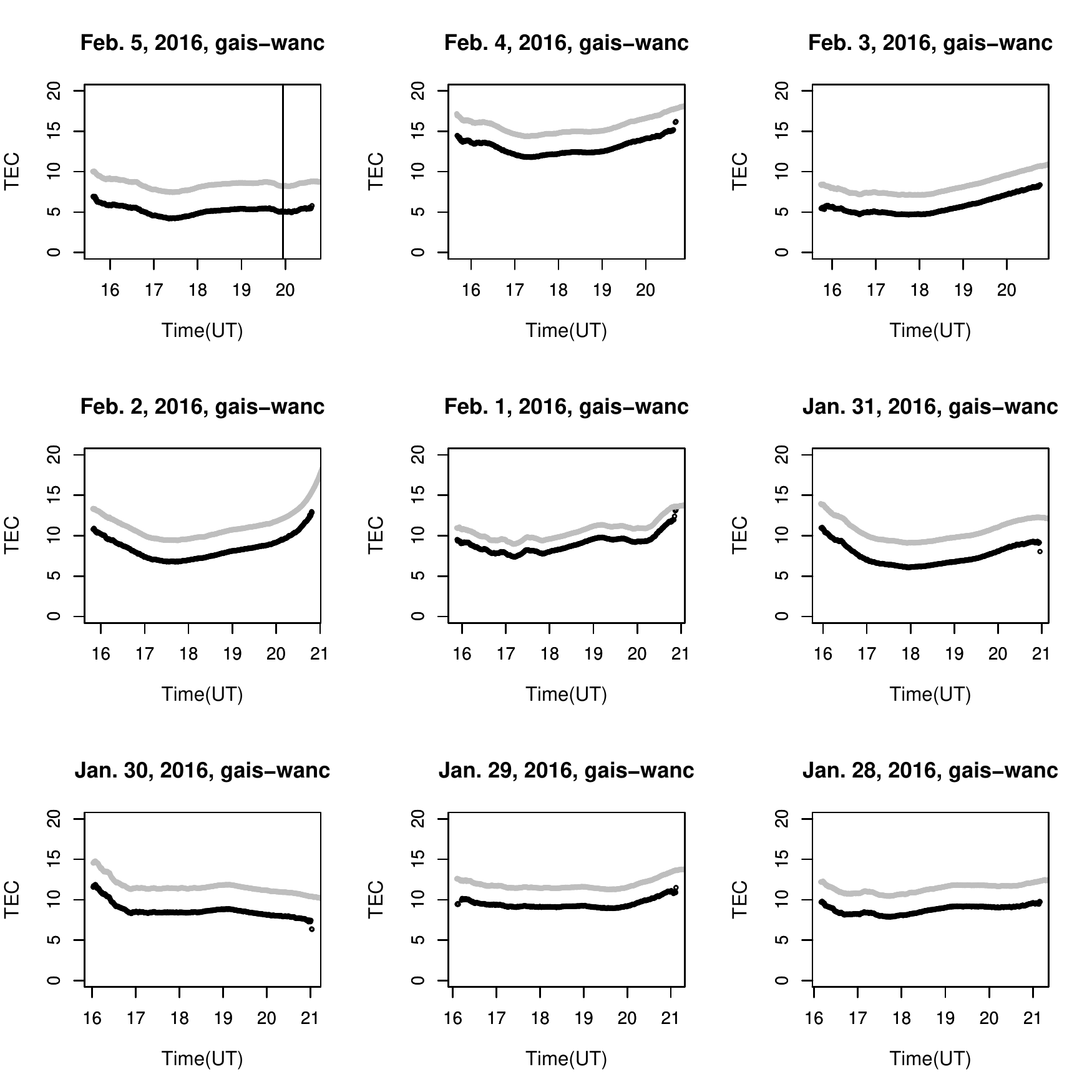}}
\caption{Time-series of TEC data observed
at  the stations ``gais'' (black) and ``wanc'' (gray) 
for the period from Jan. 28 to Feb. 5 in 2016, where the TEC values obtained
at ``wanc'' were shifted by hand for the guide of eyes.  
  GPS satellite 28 was used, and the vertical line on Feb. 5 indicates
  the time (UT) when the event occurred.
}
\label{figure-tainan-gais-wanc-raw-TEC-9-GPS28}
\end{figure}

  Figure \ref{figure-tainan-gais-wanc-raw-TEC-9-GPS28}
shows the time-series of TEC data observed for the consecutive 9 days   
including the event day. These time-series were obtained
with GPS satellite 28 associated with the stations ``gais'' and ``wanc''.
As well as the case of GPS satellite 17
( See Figure \ref{figure-tainan-gais-wanc-raw-TEC-9}),  
 it is verified that the obtained $2$ time-series on each day
were similar. 
 On the event day, time-series of TEC data were smoother than
 those of GPS satellite 17. This smoothness yields  
 low correlation values obtained with CRA.
 On Feb. 1, the time-series of TEC data were not smooth to some extent,
 and they were consistent with the case of GPS satellite 17.

\subsubsection{Correlations with $M=1$ and GPS 17 and 28
  (From Jan. 28 to Feb. 5)}

\begin{figure}[ht!]
\centerline{\includegraphics[height=15.0cm]{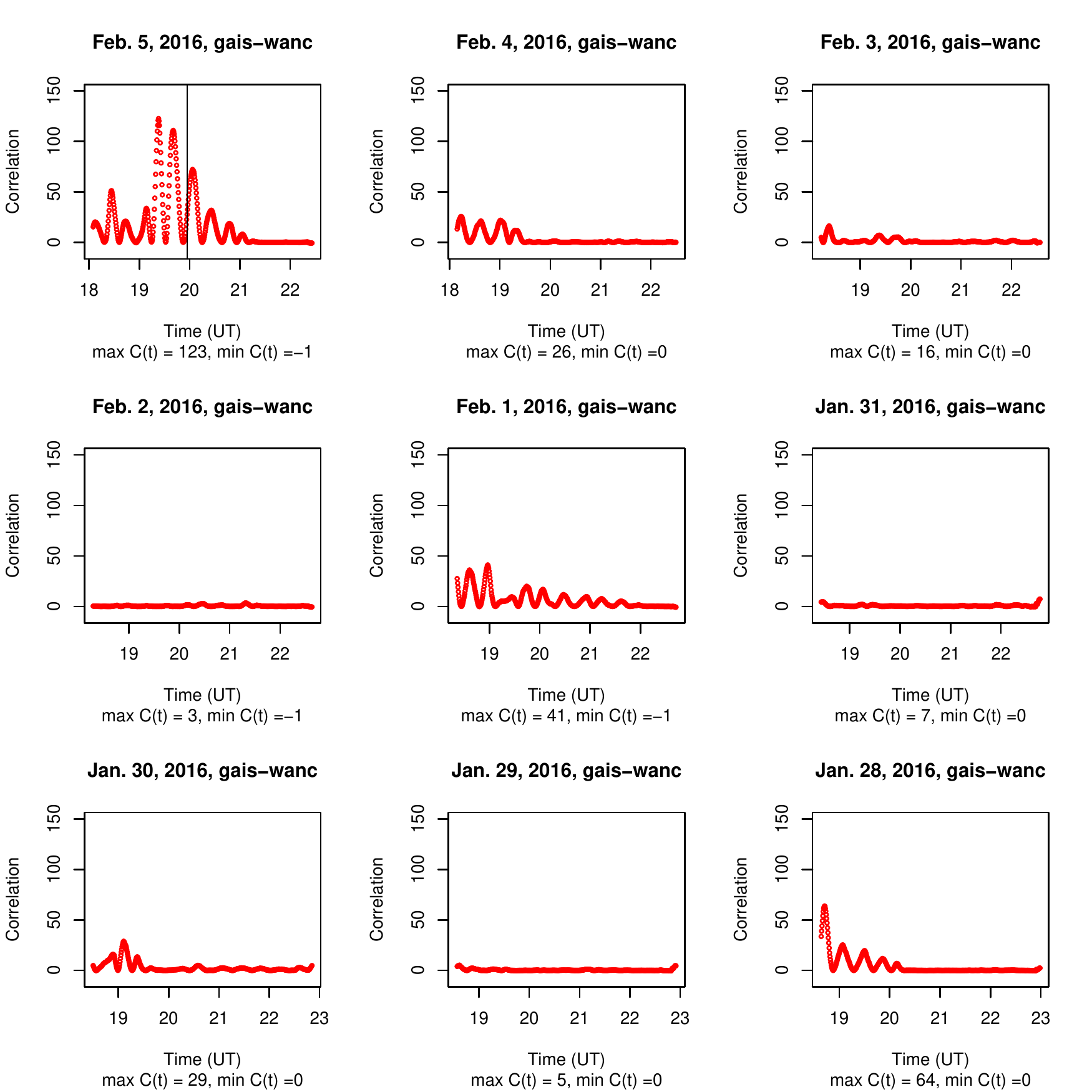}}
\caption{
Time-series of correlations obtained by CRA with $M=1$, where
  TEC data were observed 
at  the stations ``gais'' and ``wanc''
for the period from Jan. 28 to Feb. 5 in 2016.  
GPS satellite 17  was used. The vertical line on Feb. 5 indicates
the time when the event occurred.}
\label{figure-tainan-gais-wanc-correlation-9}
\end{figure}

Figure \ref{figure-tainan-gais-wanc-correlation-9} 
shows the time-series of correlations obtained by CRA
with $M=1$
and GPS satellite 17 
for the consecutive $9$ days   
including the event day.   
There were  
some periods where correlation values
are high on Jan. 28 and Feb. 1.  
On Jan. 28, this  
 anomalous period
was situated around the
time when time-series
started.  
In addition the maximum value
of the correlation on Jan. 28 
was about half of that on Feb. 5.
On Feb. 1,  
 the maximum  
value of the correlation obtained by CRA was  
$1/3$ of that on Feb. 5. 
All the time-series of correlations obtained by CRA 
  in this figure are compared with those for
  the case of $M=5$ in figure \ref{figure-tainan-gais-6-correlation-9}.

\begin{figure}[ht!]
\centerline{\includegraphics[height=15.0cm]{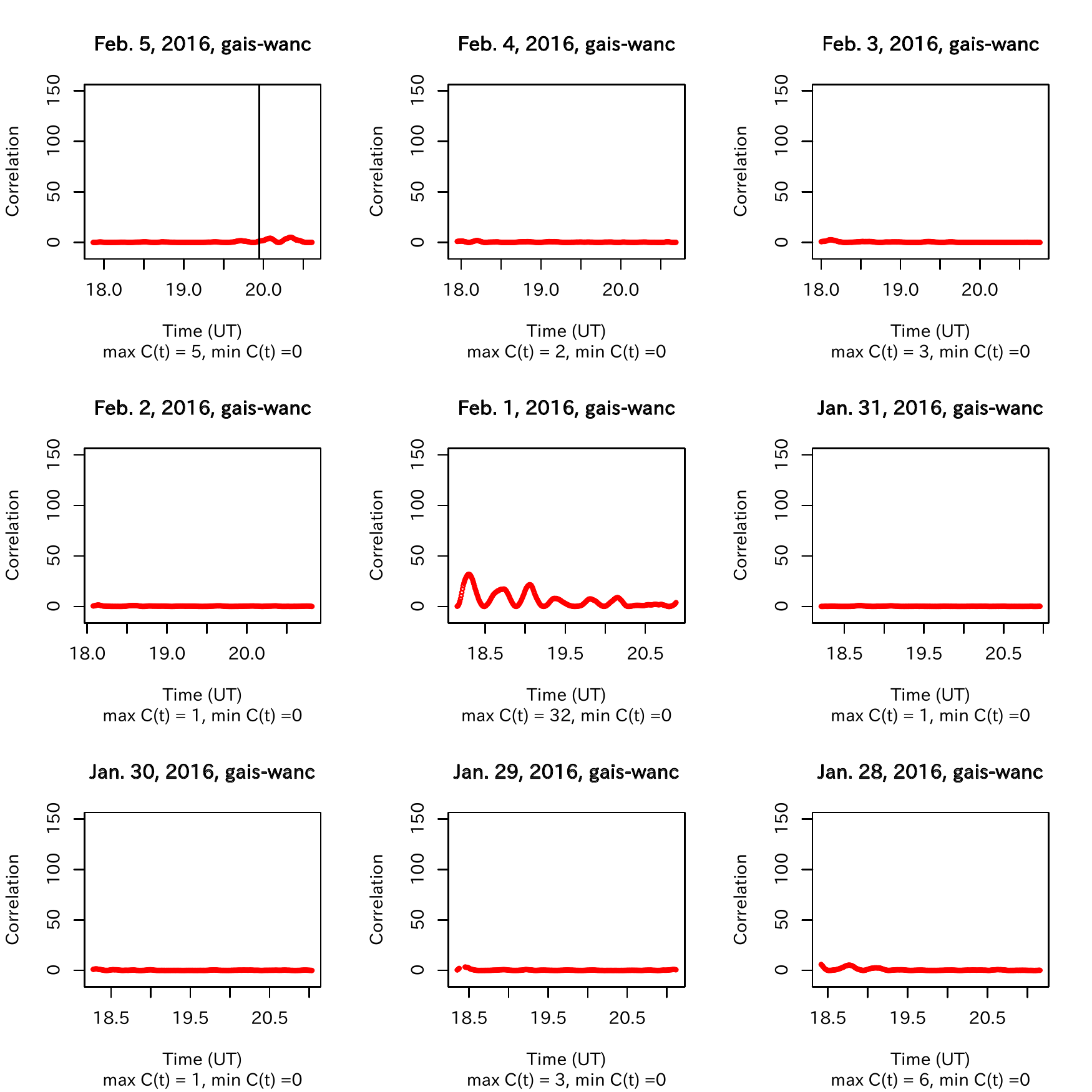}}
\caption{
  Time-series of correlations obtained by CRA with $M=1$,
where TEC data were observed  
at  the stations ``gais'' and ``wanc''
for the period from Jan. 28 to Feb. 5 in 2016.  
GPS satellite 28  was used. The vertical line on Feb. 5 indicates
the time when the event occurred.}
\label{figure-tainan-gais-wanc-correlation-9-GPS28}
\end{figure}

Figure \ref{figure-tainan-gais-wanc-correlation-9-GPS28} 
shows the time-series of correlations obtained by CRA with $M=1$
and GPS satellite 28 
for the consecutive $9$ days   
including the event day.  
There were  
some periods where correlation values
are high on Feb. 1. Note that this period also occurred on this day
for the case of GPS satellite 17.  
Although no figure is given, it can be shown that
  correlation values with GPS satellite
  28 on Feb. 1 did not decrease for the case with $M=5$. 

\subsubsection{Correlations with $M=5$ and GPS17 (From Jan. 6 to Feb. 5)}
 
\begin{figure}[ht!]
\centerline{\includegraphics[height=15.0cm]{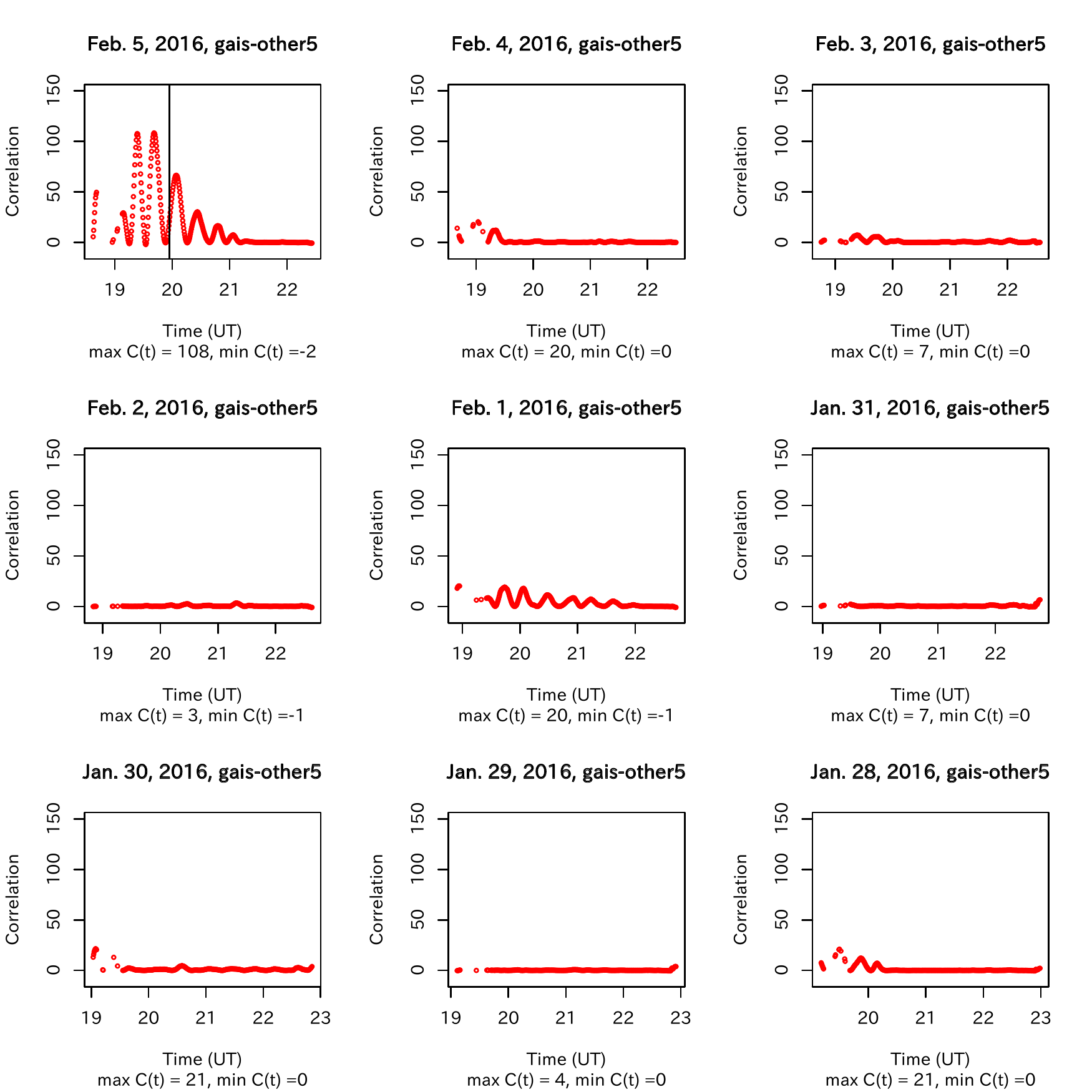}}
\caption{
Time-series of correlations obtained by CRA with $M=5$, where
  TEC data were observed 
  at 6 stations for the period from Jan. 28 to Feb. 5 in 2016. 
  GPS satellite 17  was used, and the central station
  is ``gais'', and the other 5 stations are
  ``wanc'', ``jwen'', ``mlo1'', ``dani'', ``shwa''.   
  The vertical line on Feb. 5 indicates
  the time (UT) when the event occurred.
  }
\label{figure-tainan-gais-6-correlation-9}
\end{figure}

Figure \ref{figure-tainan-gais-6-correlation-9}
shows the time-series of correlations obtained by CRA 
with $M=5$ for the consecutive 9 days   
including the event day.
The central station is ``gais'' that is the nearest one
from the epicenter. The other stations were chosen as 
``wanc'', ``jwen'', ``mlo1'', ``dani'', and ``shwa'' so that 
they are located near the epicenter.
When the number of stations for the correlation analysis increases, 
it is expected that the correlation values 
become higher 
than those with $M=1$,
if each station receives common sudden
disturbed signals.   
On the event day,
the obtained time-series is 
consistent with the case with $M=1$ 
in Figure \ref{figure-tainan-gais-wanc-correlation-9}.  
For the days Feb. 1 and Jan. 28, 
on which correlation values were high 
but no event, 
the maximum values decreased
by more than half of those 
with $M=1$.
Combining these, we see that 
the anomaly in time-series on Feb. 5 appeared commonly around the central
station ``gais''. On the contrary,  anomalies on Feb. 1 and Jan. 28 did not
appear commonly around ``gais''
for the SIP track for GPS satellite 17.

\begin{figure}[ht!]
\centerline{\includegraphics[height=15.0cm]{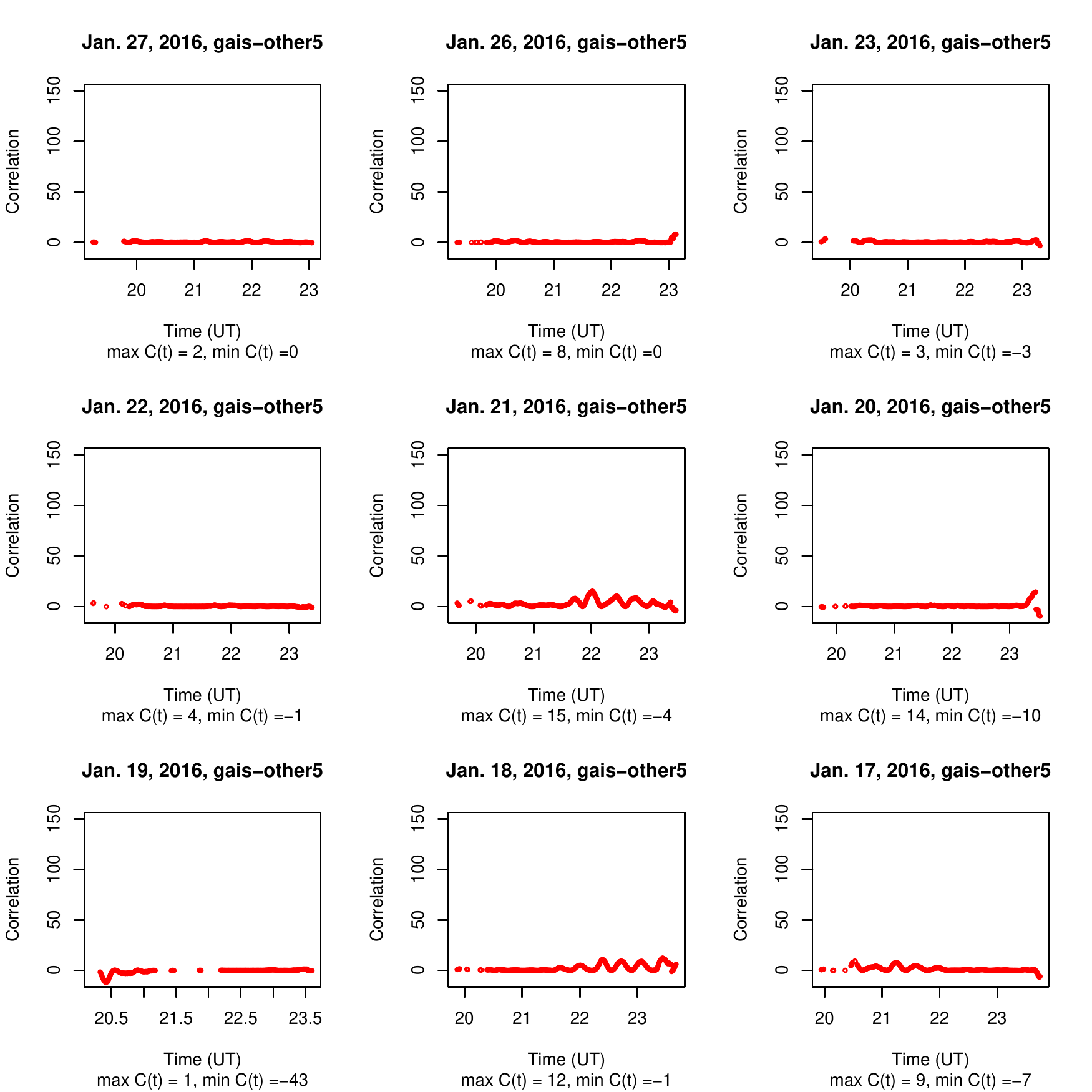}}
\caption{
  Time-series of correlations obtained by CRA with $M=5$ 
  for the period from Jan. 17 to Jan. 27 in 2016, except for Jan. 24 and 25,  
  at the stations described in the caption of
  Figure \ref{figure-tainan-gais-6-correlation-9}.
  GPS satellite 17 was used. 
}
\label{figure-tainan-gais-wanc-correlation-9-2-M5-27}
\end{figure}

Figure \ref{figure-tainan-gais-wanc-correlation-9-2-M5-27} shows 
time-series of correlations obtained by CRA with $M=5$  
for the period
from Jan. 17 to Jan. 27 in 2016, except for Jan. 24 and 25. 
  On Jan. 24 and 25, no data were acquired. 
No significant anomaly was observed in this period,  
and this observation is consistent with the
non-existence of significant geomagnetic activity 
( see Section\,\ref{section-discussion} ).

\begin{figure}[ht!]
%
\centerline{\includegraphics[height=15.0cm]{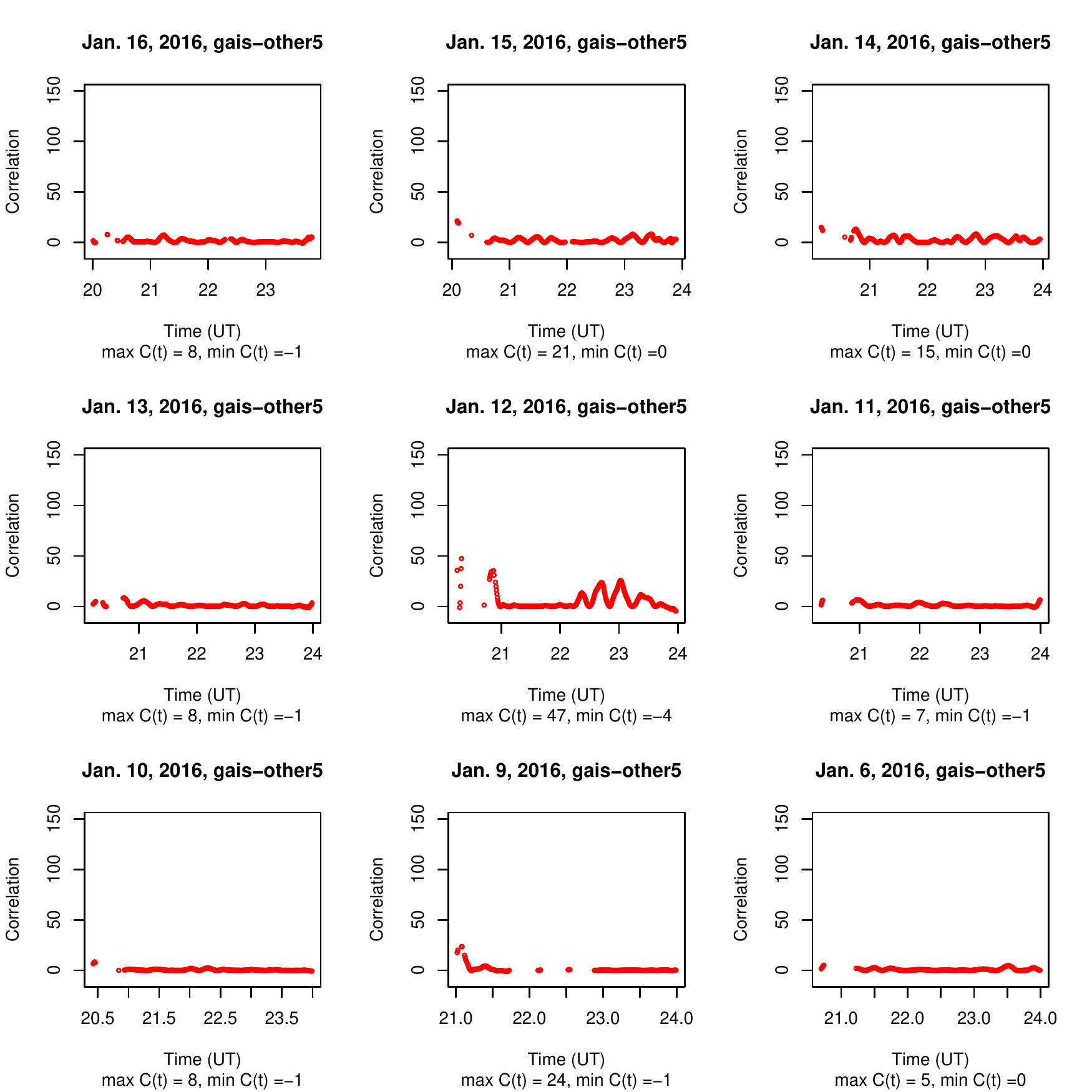}}
\caption{
  Time-series of correlations obtained by CRA with $M=5$, where 
TEC data observed at the stations described in the caption of
    figure \ref{figure-tainan-gais-6-correlation-9}
    for the period from Jan. 6 to Jan. 16 in 2016, except for Jan. 7 and 8.
    GPS satellite 17 was used. 
}
\label{figure-tainan-gais-wanc-correlation-9-2-M5-16}
\end{figure}

  Figure \,\ref{figure-tainan-gais-wanc-correlation-9-2-M5-16}
shows 
time-series of correlations obtained by CRA with $M=5$  
for the period 
from Jan. 6 to Jan. 16 in 2016, except for Jan. 7 and 8. 
On Jan. 7  and 8, no data were acquired. 
On Jan. 12th, there were some periods where
correlation values are higher 
than those on the other days in this figure.  
Although no figure is given, it can be shown that
such correlation values were lower than those with $M=1$ for GPS satellite 17.
This reduction of correlation values as $M$ increases 
is consistent with the case 
where there was no significant seismic activity around the island.

\section{Discussion}
\label{section-discussion}
TEC anomalies before the earthquake have been shown
in Section\,\ref{section-data-presentation-2016}. 
To give another evidence that such TEC anomalies were caused by the Taiwan 
earthquake, our correlation analysis can also
be applied to time-series of TEC data
obtained in Japan on the event day.
GNSS data were obtained from the 
 Geospatial Information Authority of Japan. 
The distance between Taipei and Fukuoka located the south part of
Japan is about $1280$ km, 
and the correlation analysis applied to TEC data
observed in Japan provides how much 
Medium-Scale Traveling Ionospheric Disturbance 
(MSTID) occurred, where
Japan is located upstream of MSTID\cite{Hunsucker1982,Otsuka2011}. 
Here MSTID is known as a type of disturbance in the ionosphere,
and it might hide TEC anomalies as an earthquake precursor. 

\begin{figure}[ht!]
%
%
\includegraphics[height=7.4cm]{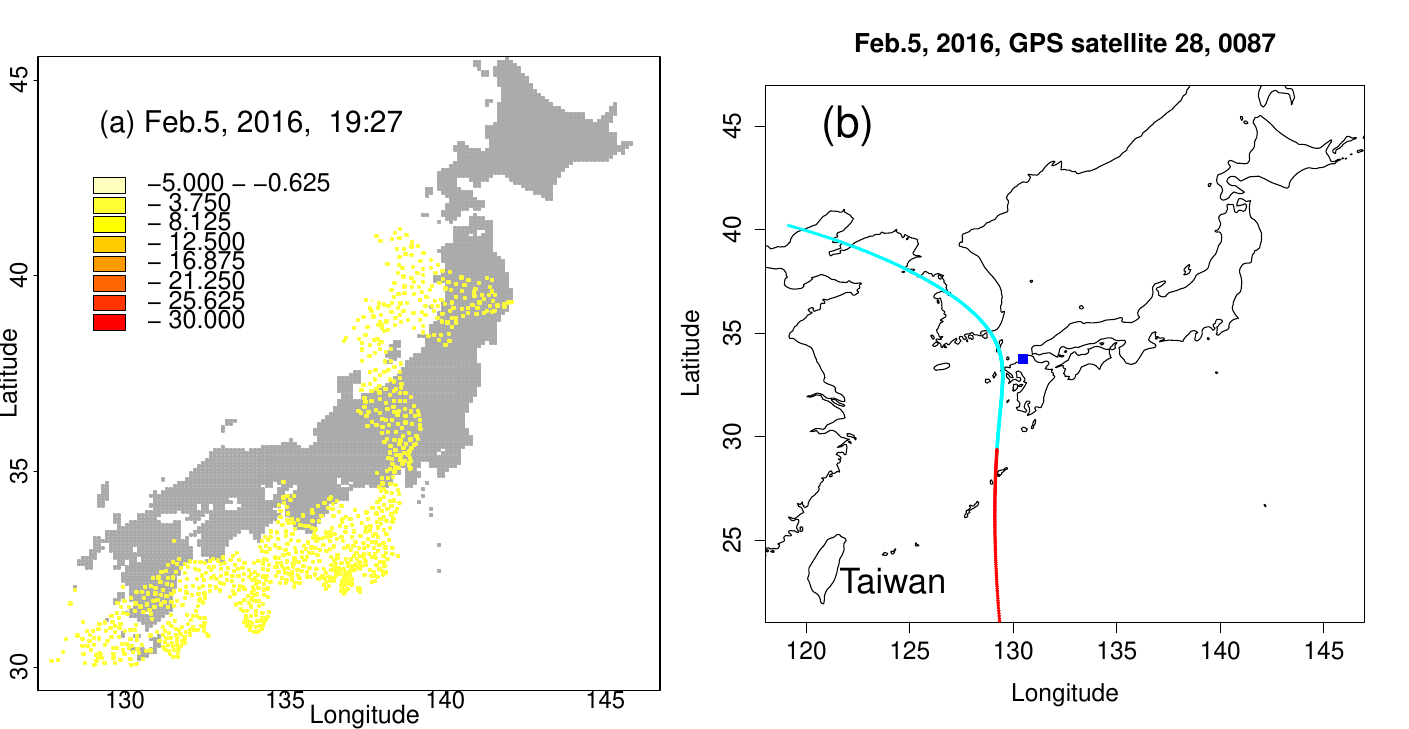}

  \caption{(a) Time-seriese of correlations obtained by CRA 
  with $M=30$, where   TEC data were
  observed
  at various stations in Japan at 19:27 UT
  on Feb. 5, 2016.
  Here GPS satellite 28 was used.
  (b) SIP track from GPS satellite 28
  for
  ``0087'' (``Koga'' in Fukuoka, blue, $\blacksquare$ ).
  Cyan curve: SIP track before the event, and
  Red curve: SIP track after the event.
}
\label{figure-Japan}
\end{figure}

Figure \ref{figure-Japan}(a) shows the time-series of correlations with $M=30$
obtained with TEC data 
observed at various GNSS stations in Japan 30 min
before the event, 
where TEC data were obtained  
with GPS satellite 28.
It suggests that there was no significant
TEC anomaly and MSTID around Japan.    
Figure \ref{figure-Japan}(b) shows the SIP track 
from GPS satellite 28 for the  
 GNSS station ``0087'', Fukuoka in Japan.
The altitude for calculating SIP  
is $300$ km above the ground.  
From  these data concerning the upstream of MSTID, there was 
no disturbance due to MSTID around the time when the anomalies in Taiwan
were observed. 

Geomagnetic variations indices of Kp and Dst are examined during
a month, from Jan. 5 to Feb. 5 2016, 
     on this earthquake day and earlier.
Provisional Dst indices show variations between $-53$ and $17$
from Jan. 23 to Feb. 5, and those between $-93$ and $19$ 
from Jan. 5 to Jan. 22. 
No Sudden Storm Commencement was observed 
 from Jan. 23 to Feb. 5,  
and then this period was under 
relatively quiet geomagnetic conditions. Therefore it is presumed 
that the analyzed period of TEC variations were
not contaminated by relatively
large geomagnetic disturbances.  
On the other hand, although some geomagnetic disturbance occurred from 
Jan. 5 to Jan. 22, it did not affect the correlation values obtained by CRA, 
since the maximum Kp index was observed around Jan. 21, and there was no 
high correlation value around Jan. 21
( See Figure \ref{figure-tainan-gais-wanc-correlation-9-2-M5-27} ).

\vspace*{1cm}

To explain TEC anomalies responsible for large earthquakes, several 
models consisting of 
electromagnetic processes due to stressed rocks   
have been proposed.
Recent progress of such has been summarized in 
\cite{Heki2017JGR}.
Such models are Lithosphere-atmosphere-ionosphere (LAI)
coupling models, and one of them is based on the experimental
results on stressed rocks \cite{Freund2013}. 
In \cite{H-Enomoto2015}, differences between
intraplate earthquakes and interplate ones with Mw$\geq 8.2$  
were argued, and then each mechanism that causes precursors  
could be different each other.   
Since the focus of this study 
is an intraplate earthquake with Mw6.4,  
the existing models proposed so far 
for observed TEC anomalies may not be
suitable.

As shown in the previous section, SIP tracks could
  be related to how TEC anomalies are observed. To explore such a 
  relation, assume that some enough amount of
  electric currents in the earth crust exist before relatively
  large earthquakes \cite{Freund2003}. From this, electromagnetic
  fields are induced in the ionosphere.  
  If an IPP track crosses an assumed ionospheric anomaly area, 
  then TEC is disturbed by the induced electromagnetic field
  ( See Figure \ref{Figures-AB} ). 
  This disturbance causes the emergence of 
  high correlation values, which is consistent with our results.
  For quantitative discussions, a physical model is needed for future study. 
  
  This assumption of electromagnetic fields
  can deny a possibility that solar activity contributed to 
  TEC anomalies on the event day.
  This is argued by showing a contradiction from an assumption.
  Assume first that an intensive solar activity contributes to TEC anomalies. 
  Then correlation values obtained by CRA with respect to each satellite 
  should equally be affected, i.e., how TEC anomalies are observed
  should not be related to SIP tracks.
  The results shown in this paper have indicated that
  it is not the case. Thus, the assumption does not hold, i.e.,
  the TEC anomalies found with CRA are not ascribed to
  the cause of solar activity.

  To refine the discussions above, we need more
  examples of this kind of analysis, physical models,  
  and some experimental verification.
  Such examples include the 2016 Kumamoto earthquake. That is also 
  an intraplate earthquake with more than Mw7.0, for which 
  CRA showed some TEC anomalies before the event
  \cite{IU2017}.

\section{Conclusions}
It has been shown how a TEC anomaly can be detected prior to the 2016 Taiwan
earthquake. 
This earthquake is with Mw6.4, and is classified as a type of
intraplate one. With CRA,   
the TEC anomalies have been detected about 40 min
before the event.
Also it has been argued that the anomalies are not ascribed to
  solar activity by showing 
  how TEC anomalies can be
  observed with respect to the SIP track. 
Since it had not been known such an hour-long earthquake precursor for
this class of moment magnitudes, this finding is a first example
for showing such a precursor. One candidate of the reasons why
this anomaly can appear is that the 2016 Taiwan earthquake is intraplate one.
There are some future works involving this study. 
One is to explore why this anomaly in the ionosphere appears with models
such as the LAI coupling model.
Another one is to apply this correlation analysis
to some intraplate earthquakes. The later
is expected to reveal conditions when this type of anomaly in the
ionosphere is observed.

%

\section*{Acknowledgments}
 The authors thank Dr. T. Yoshiki for fruitful discussions,
 the Geospatial Information Authority of Japan (GSI) for providing the GPS
 data in
Japan, the Central Weather Bureau in Taiwan for
providing the GPS data in Taiwan.
 Also, the authors belonging to Kyoto university acknowledge 
 K-Opticom cooperation for continuous support regarding the
 Kyoto University-K-Opticom collaborative research agreement (2017-2020).
C. H. Chen is supported by Central Weather Bureau (CWB) of Taiwan
to National Cheng Kung University under MOTC-CWB-107-E-01.
The GPS observations were
provided by the Central Weather Bureau (CWB)
of Taiwan.


%
%
%
%
%
%
%
%
%





\end{document}